\documentclass[twocolumn,pre]{revtex4}
\usepackage{amssymb}
\usepackage{amsmath}
\usepackage{graphicx}

\begin{document}

\title{Error--proof programmable self--assembly of DNA--nanoparticle clusters%
}
\author{Nicholas A. Licata, Alexei V. Tkachenko}
\affiliation{Department of Physics and Michigan Center for Theoretical Physics,
University of Michigan, \ 450 Church Str., Ann Arbor, Michigan 48109}

\begin{abstract}
We study theoretically a new generic scheme of programmable self-assembly of
nanoparticles into clusters of desired geometry. \ The problem is motivated
by the feasibility of highly selective DNA-mediated interactions between
colloidal particles. \ By analyzing both a simple generic model and a more
realistic description of a DNA-colloidal system, we demonstrate that it is
possible to suppress the glassy behavior of the system, and to make the
self-assembly nearly error-proof. This regime requires a combination of
stretchable interparticle linkers (e.g. sufficiently long DNA), and a soft
repulsive potential. The jamming phase diagram and the error probability are
computed for several types of clusters. The prospects for the experimental
implementation of our scheme are also discussed. \ 

\end{abstract}

\maketitle

\section{Introduction}

Over recent years, significant attention has been attracted to the
possibility of nanotechnological applications of DNA\cite{nucleic},\cite%
{falls},\cite{template},\cite{angstrom},\cite{nanotube}. \ Among the various
proposals, one of the most interesting directions is the use of
DNA--mediated interactions for \textit{programmable self--assembly} of
nanoparticle structures \cite{rational},\cite{natreview},\cite%
{designcrystals},\cite{nanocrystals}. Several schemes of such self-assembly
have been studied both experimentally and theoretically. Their common theme
is the use of colloidal particles functionalized with specially designed
ssDNA (markers), whose sequence defines\ the particle type. In such systems,
selective type-dependent interactions can be introduced either by making the
markers complementary to each other, or by using linker-DNA chains whose
ends are complementary to particular marker sequences.

Recent theoretical studies have addressed the expected phase behavior \cite%
{phasebehav}, melting properties \cite{meltingprop}, and morphological
diversity \cite{morphology} of DNA--colloidal assemblies. \ In particular,
there are indications that these techniques can be utilized for fabrication
of photonic band gap materials \cite{photonic},\cite{wiremesh}. \ Despite
significant experimental progress, the long-term potential of DNA--based
self--assembly is far from being realized. \ For instance, most of the
experimental studies of DNA--colloidal systems report random aggregation of
the particles \cite{colloidalgold}. \ Some degree of structural control in
these systems has been achieved, mostly by varying the relative sizes of
particles, rather than by tuning the interactions \cite{micelle},\cite%
{blocks}.

In the present paper, we take a broader view of programmable self-assembly.
\ While this theoretical study is strongly motivated by the prospects of
DNA-colloidal systems, our main objective is to address a more general
question: how well can a desired structure be encoded by tunable
interactions between its constituents? The particular model system on which
we focus consists of distinguishable particles with individually controlled
interactions between any pair of them. In the first stage of our study, we
analyze a simplified yet generic version of such a system. \ All the
particles have the same repulsive potential, while the attraction is
introduced only between selected pairs of particles in the form of a
spring-like quadratic potential. \ This potential mimics the effect of a
stretchable DNA molecule whose ends can selectively adsorb to the particular
pair of particles. \ We then introduce a number of additional features which
make the model a more realistic description of an actual DNA-colloidal
system.

Our major result is that a combination of stretchable interparticle linkers
(e.g. sufficiently long DNA), and a soft repulsive potential greatly reduces
(or totally eliminates) the probability of self-assembling an undesired
structure. \ The experimental prototype is a system of particles in a
mixture of two types of DNA\ molecules which can selectively adsorb to the
particle surface. \ The first type are DNA molecules with two "sticky" ends,
i.e. both end sequences of the DNA are complementary to the particle marker
sequence. \ With one end adsorbed to the particle surface, the remaining
sticky end makes it possible to introduce an attractive interparticle
potential between selected particle pairs. \ The second type are DNA\
molecules with one sticky end which adsorbs to the particle surface. \ These
DNA strands give rise to a soft repulsive potential of entropic origin
between all particle pairs. \ 

There is a natural analogy between our problem and the folding of proteins
where interactions between amino acids encode the overall structure. \
However, it should be emphasized that the task of programmable self-assembly
in colloidal systems is even more demanding than protein folding: any
intermediate metastable configuration has a much longer lifetime and
therefore means a misfolding event. \ Because of this, we were looking for a
self--assembly scenario which does not require thermally activated escape
from a metastable configuration. This makes the problem additionally
interesting from the theoretical perspective of "jamming", a phenomenon
actively studied in the context of granular and colloidal systems. Our
results can be interpreted in terms of a jamming-unjamming transition
controlled by the interaction parameters.

\ It should be emphasized that the goal of this work is primarily
conceptual, as opposed to providing a manual for the immediate experimental
realization of ordered colloidal structures. \ Nevertheless, future
experimental schemes will be forced to overcome obstacles presented by
colloidal jamming. \ With this in mind, one of the most salient features of
our model is the ability to smooth the energy landscape by tuning the
interactions between particles. \ 

The plan for the paper is as follows. \ In section II, we address the
problem within a simplified generic model which mimics the nanoparticle
system with stretchable DNA connections. An unexpected and very encouraging
result of this study is that the misfolding (or jamming) in the model system
can be completely avoided for a certain set of parameters. \ In section III
the original model is adapted to a more realistic situation which
incorporates the random character of the DNA-mediated interactions. \ In
section IV we discuss the prospects for the future experimental
implementation of our scheme. \ In section V we summarize the major results.
\ \ 

\section{Beads and Springs Model}

Consider an isolated group of repulsive particles linked via a polymer
spring to their desired nearest neighbours. \ We assume that the DNA\ marker
sequence is the same for any two markers attached to the same particle, but
different particles have different marker sequences (i.e. each particle has
a unique code). \ In this case, the attraction between any two particles can
be effectively switched on by adding DNA "linkers" whose ends are
complementary to the corresponding marker sequences of the particles. \ As a
first approach to the problem, we introduce a generic "Beads \& Springs"
model which incorporates essential features of the DNA--nanoparticle system.
\ The model system contains $N$ particles with pairwise (type-independent)
repulsive potential $U(r)$. \ In general, this repulsion may have a
hard-core or soft-core behavior, or be a combination of the two. In order to
model the DNA--induced type--dependent attraction, we introduce a harmonic
potential (linear springs) which acts only between selected pairs of
particles\cite{scmbook}. \ Thus, the model Hamiltonian has the following
form: 
\begin{equation}
H=\frac{1}{2}\sum\limits_{\alpha ,\beta }\kappa J_{\alpha \beta }\left\vert 
\mathbf{r}_{\alpha }-\mathbf{r}_{\beta }\right\vert ^{2}+U(|\mathbf{r}%
_{\alpha }-\mathbf{r}_{\beta }|)
\end{equation}
Here $\alpha ,\beta $ are the particle indices, $\mathbf{r}_{\alpha }$ are
their current positions, and $\kappa $ is the spring constant. \ The
connectivity matrix element $J_{\alpha \beta }$ may be either 1 or 0,
depending on whether the two particles are connected by a spring, or not. \
Our goal is to program\ the desired spatial configuration by choosing an
appropriate connectivity matrix $J_{\alpha \beta }$. A natural construction
is to \textit{put }$J_{\alpha \beta }=1$\textit{\ for any pair of particles
which must be nearest neighbors} in the desired cluster, and not to connect
the particles otherwise (i.e. put $J_{\alpha \beta }=0$). \ This
construction assures that the target configuration is the ground state of
the system. \ 

Note that our problem is somewhat similar to that of heteropolymer folding.
\ In that case, the selective interactions between monomers (e.g. amino
acids in protein) are responsible for the coding of the spacial structure of
a globule. Our major concern is whether the kinetics of the system will
allow\ it to reach the ground state within a reasonable time. \ \
Unfortunately, since the Brownian motion of a typical nanoparticle is
relatively slow (compared to molecular time scales), it is unrealistic to
expect that our system will be able to find the target configuration by
"hopping" between various metastable states, as in the case of protein
folding. \ However, our case is different because the attractive force grows
with distance, as opposed to the short--range nature of heteropolymer
self--interactions. \ As we shall see below, this difference is essential,
making it possible for the system to reach the ground state without stopping
at any metastable configuration.

\begin{figure}[h]
\includegraphics[width=3.2in,height=3.47in]{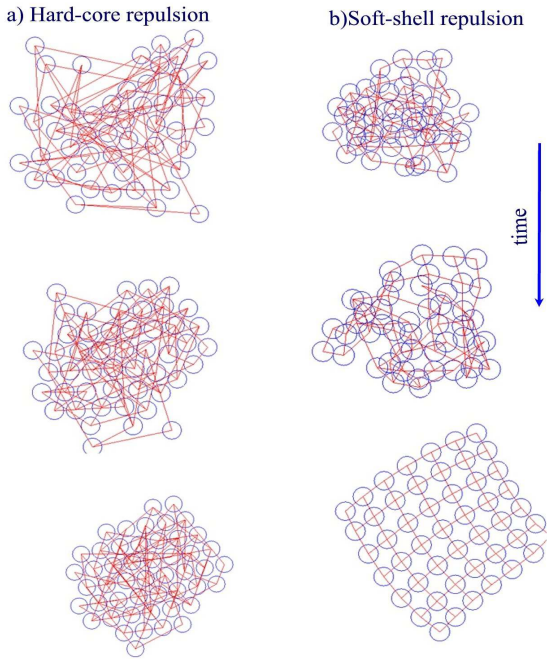}
\caption{(Color Online). Programmable self--assembly in 2D, studied within
the Beads \& Springs model. 49 particles are all distinct and connected with
linear springs to encode the desired configuration (7x7 square). The jamming
behavior is observed for the case of hard spheres (a). However, the assembly
of the target structure can be achieved if the repulsion is sufficiently
"soft" (b). }
\label{2D}
\end{figure}

We have performed a molecular dynamics simulation of the above model by
numerically integrating its Langevin equation:

\begin{equation}
b^{-1}\mathbf{\dot{r}}_{\alpha }=-\mathbf{\nabla }_{\alpha }H+\mathbf{\eta }%
_{\alpha }
\end{equation}
Here $b$ is the particle mobility. \ The thermal noise has been artificially
suppressed in this study (i.e. $\eta =0$ ). \ In other words, we have
assumed the worst case scenario: once the system is trapped in a local
energy minimum, it stays there indefinitely. The equations of motion were
solved numerically by a first order Runge-Kutta method. \ First, we studied
a system of $N=49$ distinguishable particles in 2D, whose native
configuration was a $7\times 7$ square cluster (see Figure \ref{2D}). \
Their initial positions were random, and the connectivity matrix was
constructed according to the above nearest--neighbor rule. \ 

First, we studied the case when $U(r)$ is a hard--core potential. More
precisely, the repulsive force was given by a semi--linear form: 
\begin{equation}
f_{hc}\left( r\right) =-\frac{\partial U_{hc}\left( r\right) }{\partial r}%
=\kappa _{0}(d-r)\Theta (d-r)
\end{equation}
Here $\Theta $\ is the unit step function, and $d$ is the diameter of the
hard sphere. \ The parameter\ $\kappa _{0}$ determines the strength of the
hard-core repulsion, and it does not affect the results, as long as $\kappa
_{0}\gg \kappa $. \ In our simulations, we found that the hard sphere system
eventually stops in a configuration definitely different from the desired
one, a behavior which is well known in the context of granular and colloidal
systems as "jamming"\cite{geltransition}. \ Remarkably, \textit{the jamming
can be avoided when the hard-core repulsion is replaced by a soft--core
potential}: 
\begin{equation}
U_{sc}(r)=U_{0}\exp (-r/\lambda ).
\end{equation}
Here the decay length $\lambda $ is of the order of the equilibrium
interparticle distance $r_{0}$. \ This indicates that the energy landscape
can be made smooth by a combination of long-range selective attraction and
soft-core repulsion. \ 

The result is surprising and remarkably robust. \ In particular, in order to
expand our finding to the 3D case, we studied the self-assembly of particles
into tetrahedra of various sizes ($N=10,20,35$). \ This time, the hard core
interaction potential was superimposed with\ a soft shell repulsion, which
makes the model more relevant for an actual DNA-colloidal system: 
\begin{equation}
U\left( r\right) =U_{hc}\left( r\right) +U_{sc}\left( r\right) .
\end{equation}%
After the system has fully relaxed, a geometric measure of the folding
success is determined by comparing particle separations of the desired final
state to those generated from a set of random initial conditions. Figure \ref%
{jammed} shows the "jamming phase diagram" for these systems. \ To assign a
point on the diagram to the correct folding regime, we required 100
consecutive successful folds. \ While this criterion can only give an upper
bound on the jamming probability (which is approximately $1\%$), an
additional analysis gives strong evidence that the correct folding region of
the diagram corresponds to zero probability of jamming. \

\begin{figure}[h]
\includegraphics[width=3.2776in, height=2.4699in]{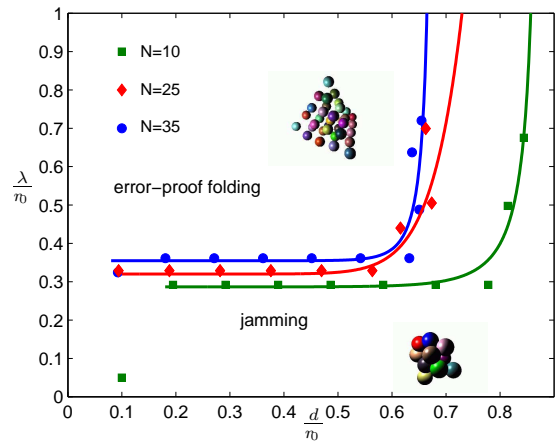}
\caption{ (Color Online). "Jamming phase diagram"
obtained for programmable self-assembly of tetrahedral clusters, within the
Beads \& Springs model. The control parameters depend on the equilibrium
interparticle distance $r_{o}$, the diameter of the hard sphere $d$, and the
range of the soft-shell repulsion $\protect\lambda $.}
\label{jammed}
\end{figure}

\section{Self-Assembly in DNA colloidal systems}

As we have seen, the introduction of a soft-core repulsive potential $U_{sc}$
is crucial to a successful self-assembly proposal. \ In a real system, this
repulsion can be generated e.g. by DNA\ or another water soluble polymer
adsorbed to the particle surface. \ The mechanism is quite independent of
the monomer chemistry, but for the sake of concreteness we will speak of the
repulsion generated by DNA. Namely, we assume that a certain fraction of the
DNA "arms" of the "octopus-like" particles are not terminated by a sticky
end, and only play the role of a repulsive "buffer". \ When the polymer
coverage is sufficiently low, the interparticle repulsion is primarily due
to entropy loss of a chain squeezed between two particles. \ The
characteristic length scale of this interaction is given by the radius of
gyration of the "buffer" chain, $R_{g}$. \ The corresponding repulsive force 
$f_{sc}$ can be calculated exactly in the limit of relatively short buffer
chains, $R_{g}\ll d$. The result of this calculation\cite{morphology} can be
adequately expressed in the following compact form: \ 
\begin{equation}
f_{sc}\left( r\right) \approx \frac{4NR_{g}k_{B}T}{d\left( r-d\right) }\exp
\left( -\frac{\left( r-d\right) ^{2}}{2R_{g}^{2}}\right)
\end{equation}
Here $N$ is the total number of buffer chains per particle. Note that even
though this result is only valid for $R_{g}\ll d$, it correctly captures the
Gaussian decay of the repulsive force expected for longer chains as well.
Therefore, we expect the results to be at least qualitatively correct beyond
the regime of short buffer chains.\ 

In addition to the modified soft potential, we have to take into account the
random character of the realistic DNA-mediated attraction. \ It originates
from the fact that (1) the number of DNA "arms" of the original octopus-like
particles will typically be determined by a random adsorption process, and
(2) the fraction of the DNA chains recruited for linking a particular pair
of particles is also random. \ In terms of our original model, this means
that the "springs" will not have the same spring constant. \ If the
individual linkers are modelled by Gaussian chains \cite{statprop}, the
overall spring constant for a \ particular pair of connected particles is
given by $\kappa _{\alpha \beta }=k_{B}Tm_{\alpha \beta }/2R_{g}^{^{\prime
}2}$ , where $R_{g}^{^{\prime }}$ is the\ radius of gyration of a single
linker, and $m_{\alpha \beta }$ is the number of individual chains
connecting the particles. \ We assume that this number obeys the generic
Poisson distribution: \ $P(m)=\overline{m}^{m}e^{-\overline{m}}/m!$. \ As
formulated, the model is cast as a system of coupled differential equations:
\ 
\begin{equation}
\mathbf{\dot{r}}_{\alpha }=b\sum\limits_{\beta }\left[ -\frac{k_{B}T}{%
2R_{g}^{^{\prime }2}}J_{\alpha \beta }m_{\alpha \beta }r_{\alpha \beta
}+f_{hc}\left( r_{\alpha \beta }\right) +f_{sc}\left( r_{\alpha \beta
}\right) \right] \mathbf{\hat{n}}_{\alpha \beta }
\end{equation}
Here $r_{\alpha \beta }=\left\vert \mathbf{r}_{\alpha }-\mathbf{r}_{\beta
}\right\vert $, $\mathbf{\hat{n}}_{\alpha \beta }=\left( \mathbf{r}_{\alpha
}-\mathbf{r}_{\beta }\right) /r_{\alpha \beta }$.

\begin{figure}[h]
\includegraphics[width=3in, height=2.2in]{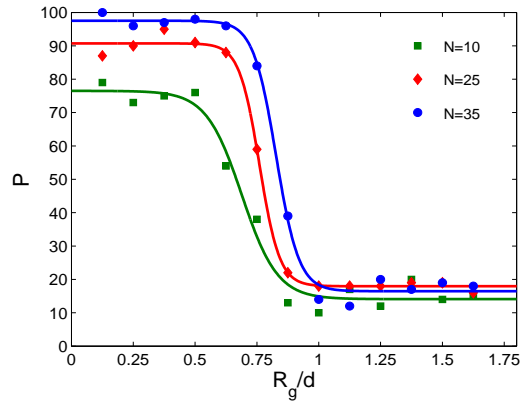}
\caption{(Color Online). Error probability $P$ as a function of aspect ratio 
$R_{g}/d$ for tetrahedral clusters with modified soft-potential and
realistic DNA-mediated attraction. Each data point on the misfolding profile
represents 100 trials. \ }
\label{error}
\end{figure}

We have studied the behavior of the system as a function of two
dimensionless parameters, one of which is the ratio of the buffer radius of
gyration to the particle diameter, $R_{g}/d$ . The other parameter
characterizes the relative strength of the attractive and repulsive forces: 
\begin{equation}
\alpha =\frac{\overline{m}}{N}\left( \frac{R_{g}}{R_{g}^{^{\prime }}}\right)
\end{equation}
Including the above modifications of the model, the essential result of the
study is that the jamming probability can be drastically suppressed,
similarly to the previous case. However, the jamming in this system cannot
be eliminated completely. \ Instead of an actual "phase boundary", we have
observed a sharp crossover to the regime of predominantly good folding, in
which the error probability is suppressed to a modest level $\sim 10-20\%$.
Interestingly, the behavior is nearly independent of the energy parameter $%
\alpha $, which makes $R_{g}/d$ $\ $\ the only major control parameter.
Figure \ref{error} shows the error probability as a function of this
geometric parameter for tetrahedral clusters of different sizes. As this
plot indicates, the misfolding behavior gets suppressed as $R_{g}/d$ exceeds 
$1$.

\ 

\begin{figure}[h]
\includegraphics[width=3in,height=2.2in]{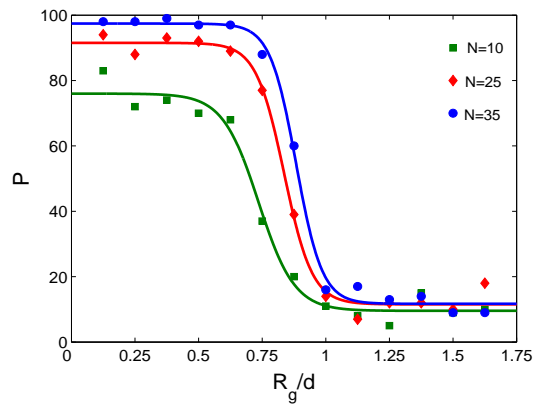}
\caption{(Color Online). Error probability $P$ as a function of aspect ratio 
$R_{g}/d$ for tetrahedral clusters in the Pincus regime. \ Each data point
on the misfolding profile represents 100 trials. \ }
\label{pincus}
\end{figure}

As a further test of the robustness of the model, we consider a modified
attractive potential which deviates from the linear Hooke's law. \ For
larger forces we enter the Pincus regime\cite{pincus}, where the end-to-end
extension of the polymer chain $r$ is related to the external tension $f\sim 
$ $r^{\frac{3}{2}}$. \ This tension law incorporates the excluded volume
interaction of individual linker DNA with themselves, which was not
previously considered. \ Remarkably, the major result for error suppression
carries over, as illustrated in figure \ref{pincus}. \ \ 

Our description has several limitations which are to be addressed in future
work. \ \ In particular, our discussion is only applicable to the limit of
modest coverage of particles with buffer chains. This case of weakly
overlapping adsorbed chains (known as the "mushroom regime") is drastically
different from the high--coverage "polymer brush" behavior\cite{scaling}. \
\ Nevertheless, our major conclusions appear to be rather robust. \ The
condition for error suppression($R_{g}/d\gtrsim 1$) is the same in both the
harmonic and Pincus regimes. \ 

\section{Discussion}

Previous studies have demonstrated an experimental implementation of
self--assembly in a DNA--colloidal system\cite{chaikin},\cite{crocker}. \
The approach in these studies differs from our vision of controllable
self-assembly in two major ways: (1) because the "linker" DNA\ chains($\sim
20nm$) used are much shorter than the particle diameter($\sim 1\mu m$), the
particles behave as sticky spheres, and (2) there is little diversity in
particle type, where the structures result from interactions of a one or two
component system. \ We would like to discuss issues related to a future
experimental implementation similar to the modified Beads \& Springs model.
\ 

In our molecular dynamics simulations we assumed that (1) the desired group
of particles has already been localized in a small region of space, and that
(2) the interparticle connections have already been made. \ There are a
number of experimental challenges associated with implementing the
self-assembly proposal of our simulations. \ In another manuscript\cite%
{packings} we provide a detailed discussion of the localization problem,
which is the first major experimental intermediary. \ 

After localization, the next step is to make the desired connections between
particles within the cluster. \ To do so one can add short ssDNA with
sequences $\bar{s}_{A}\bar{s}_{B}$ to link particles $A$ and $B$. \ The DNA
marker sequence $s_{A}$ for particle $A$ is a sequence of nucleotides
complementary to the $\bar{s}_{A}$ portion of the linker sequence $\bar{s}%
_{A}\bar{s}_{B}$. \ The hydrogen bonding of complementary nucleotides forms
base pairs which join both marker strands to the linker, creating a DNA\
bridge between the two particles. \ After the interparticle links are
formed, they should be made permanent by ligation. \ Since the spring
constants of the above dsDNA chains are too small to drive the self-assembly
of a desired cluster, we propose to melt them either by changing the
temperature or pH. As a result, the dsDNA links will be turned into ssDNA
with a much higher effective spring constant (due to the shorter persistence
length). \ This will trigger the self--assembly scenario similar to the one
discussed within the Beads \& Springs model. \ Note that DNA entanglements
may be effectively eliminated if the procedure is done in the presence of
DNA Topoisomerases. \ 

\section{Conclusion}

In conclusion, we presented a model of DNA-colloidal self-assembly which
exhibits a tunable jamming-unjamming transition. \ The combination of a
soft-core repulsion with a type-dependent long range attraction provides a
natural funneling of the energy landscape to the ground state configuration.
\ This is to be contrasted with the case of protein folding, where under
physiological conditions the interactions between amino acids are screened
to several angstroms. \ Because this lengthscale is much shorter than the
spatial extent of the native structure, large regions of the energy
landscape are flat, which prohibits formation of the native state on the
basis of funneling alone. \ As a result the folding rate is necessarily
limited by the diffusion of amino acid segments looking for their desired
nearest neighbors \cite{proteinlandscape},\cite{proteinfolding}. \ The fact
that the potential in the DNA-colloidal system is long ranged is essential,
allowing us to avoid the pitfall of slow particle diffusion. \ 

Within the Beads and Springs Model, we obtained the jamming phase diagram
for several modest sized tetrahedral clusters. \ We identified a regime of
parameter space with error-proof folding, and demonstrated the importance of
introducing a soft-core repulsion. \ The original model was then adapted to
include several features of realistic DNA-mediated interactions. \ Although
the jamming cannot be completely eliminated in the modified system, we
identified a regime of predominantly good folding, and calculated the error
probability for tetrahedral clusters. \ The jamming behavior is determined
by a single geometric parameter $R_{g}/d$. \ We concluded by discussing
prospects for an experimental implementation of our self-assembly scheme. \ 

\section{Acknowledgements}

This work was supported by the ACS Petroleum Research Fund (grant PRF
\#44181-AC10), and by the Michigan Center for Theoretical Physics (award \#
MCTP 06-03). \ 

\bibliographystyle{achemso}
\bibliography{acompat,dna}

\end{document}